\documentstyle[multicol,aps,epsf]{revtex}
\begin{document}
\title{Domain Statistics in Coarsening Systems}
\author{P.~L.~Krapivsky$\dag$ and E.~Ben-Naim$\ddag$}
\address{$\dag$Center for Polymer Studies and Department of Physics,
Boston University, Boston, MA 02215}
\address{$\ddag$Theoretical Division and Center for Nonlinear Studies, 
Los Alamos National Laboratory, Los Alamos, NM 87545}
\maketitle
\begin{abstract}
  We study the domain number and size distributions in the
  one-dimensional Ising and $q$-state Potts models subject to
  zero-temperature Glauber dynamics. The survival probability of a
  domain, $S(t)\sim t^{-\psi}$, and an unreacted domain,
  $Q_1(t)\sim t^{-\delta}$, are characterized by two independent
  nonrtrivial exponents. For the Ising case, we find $\psi=0.126$ and
  $\delta=1.27$ using numerical simulations.  We develop an
  independent interval approximation (IIA) that predicts the
  qualitative behavior of the domain distribution and provides good
  estimates for the exponents. Exact results for the domain
  distribution are also obtained in several solvable cases.

\end{abstract}

\begin{multicols}{2}

\section{INTRODUCTION} 

When a system is quenched from a homogeneous high-temperature
disordered state into a low-temperature state it does not order
instantaneously; instead, domains of equilibrium ordered phases form
on larger and larger scales\cite{bray}. It has been generally
confirmed that a scale-invariant morphology is developed, i.e. the
network of domains is (statistically) independent of time when lengths
are rescaled by a single characteristic length scale that typically
grows algebraically with time.  However, even for simple coarsening
processes little is known about more subtle properties such as the
domain size distribution\cite{ab,dz}.  One such feature that has
attracted considerable interest recently concerns the ``persistence''
of the local order parameter, the probability that it has not changed
sign in a given time interval.  Persistence has been investigated
theoretically\cite{dbg1,dbg2,dhp,bfk,ms,mbcs} and
experimentally\cite{ypms} in spin systems, interacting particles
systems\cite{kbr,cardy,krl,bhm}, Lotka-Volterra models \cite{lpe,fk},
breath figures growth\cite{bou}, foams\cite{ld}, and even simple 
diffusion \cite{msbc,w}.

Single spin persistence provides a natural counterpart to the survival
probability in the realm of many particle systems. In the context of
reaction processes, persistence is equivalent to the survival of
immobile impurities and therefore does not provide information about
collective properties of the bulk.  In contrast, domains are the
natural spatial elements of a coarsening process.  In this paper, we
ask for example what is the survival of an entire domain, $S(t)$?  This
quantity decreases with time as a power-law $S(t)\sim t^{-\psi}$.
Similar to other critical exponents, $\psi$ is universal in the sense 
that it is independent of many details such as the initial conditions.
However, it is model dependent and in this respect differs from the
growth exponent that depends only on the conservative nature of the
dynamics.

In the present work, we examine systems with short-range interactions
described by a scalar non-conserved order parameter. We focus on the
1D Ising model and its generalization to the $q$-state Potts
model\cite{wu} evolving according to Glauber spin-flip
dynamics\cite{glauber}.  Sequential dynamics has been chosen without
loss of generality as parallel dynamics exhibit similar asymptotic
behavior\cite{privman}.  In one-dimensional systems with short-range
interactions, the order-disorder transition takes place at $T=0$
\cite{dyson}, and since we are interested in coarsening, we restrict
attention to zero temperature.

This paper is organized as follows. In the next section, we discuss
the Ising and Potts models.  We define the domain number distribution
and determine it analytically in the limiting cases, $q\to\infty$ and
$q\to 1$.  Otherwise, we develop an Independent Interval Approximation
(IIA) that assumes no correlations between adjacent domains.  The IIA
predictions compare well with Monte Carlo simulations by giving
correct description of the domain statistics as well as good estimates
for the underlying exponents.  Section II is concluded with the
$q=\infty$ Potts model in arbitrary dimension $d\geq 2$.  In Sec.
III, we obtain the domain distribution in two solvable cases: the
Potts model with only energy lowering transitions and the
deterministic ballistic annihilation model \cite{ef,lpe,fk}.  Section
IV discusses some open issues and contains a summary.

\section{ISING AND POTTS MODELS}

\subsection{The Ising model and the $q$-state Potts model}

We start with the 1D Ising model subject to $T=0$ Glauber dynamics
\cite{glauber}. To examine the role of the number of equilibrium
phases we also consider a generalization of the Ising model, the
$q$-state Potts model.  Experimental realizations are known for $q=2$
(the Ising model) and $q=3,4,\infty$\cite{wu}.  The $q=\infty$ case
describes several cellular structures\cite{st}, e.g.,
polycrystals\cite{poly}, foams\cite{ld}, soap froth\cite{glaz}, and
magnetic bubbles\cite{bubbl}.

We consider uncorrelated initial conditions where each of the $q$
phases is present with equal density $1/q$.  The $T=0$ Glauber-Potts
dynamics proceeds by selecting a spin at random and changing its value
to that of one of its randomly selected neighbors.  Thus, domain walls
perform a random walk and upon contact, they annihilate or coalesce,
depending on the state of the corresponding domains
\cite{bbd,racz,amar}. Identifying a domain wall with a particle,
($A$), and absence of a domain wall with a hole ($0$), one finds the
single-species diffusion-reaction process \cite{bbd,racz,amar}

\begin{eqnarray}
\label{aa}
A0&\buildrel {1\over 2} \over \rightleftharpoons &0A,\nonumber\\
AA&\buildrel {1\over q-1} \over \longrightarrow &00,\\
AA&\buildrel {q-2\over q-1}\over \longrightarrow&A0\ {\rm or }\ 0A. 
\nonumber
\end{eqnarray}
The rates indicate the relative probabilities by which each event 
occurs.

\subsection{Domain Number Distribution: Definition and Scaling Properties}

Our goal in this study is to investigate $S(t)$, the probability that
a domain, initially present at the system at time $t=0$, has not
flipped up to time $t$ (see Fig.~1). We will present theoretical and
numerical evidence supporting an algebraic long time decay of this
survival probability,

\begin{equation}
\label{st}
S(t)\sim t^{-\psi}. 
\end{equation}
Such a behavior is robust, as the exponent $\psi$ is not sensitive to
initial state of the system (provided long ranged correlations are
absent). Our results will also strongly suggest that the exponent
$\psi$ is nontrivial, i.e., it cannot be extracted from so-far known
exponents associated with the Ising model.

\begin{figure}
\vspace{-.25in}
\centerline{\epsfxsize=7cm \epsfbox{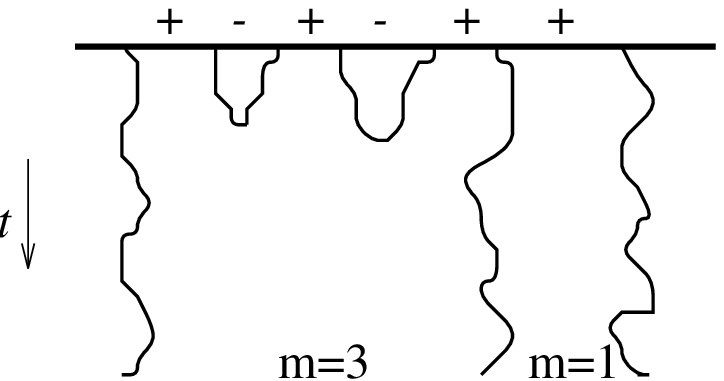}}
\noindent
{\small {\bf Fig.~1}.  
Domain motion in the Ising model. Surviving domains 
are marked by $+$, annihilated domains by $-$. The domain number 
at a later time is also indicated.}
\end{figure}

In principle, a surviving domain may undergo coalescence with other
similar phase domains.  Thus, a natural generalization of the domain
survival probability is $Q_m(t)$, the density of domains composed of
$m$ original domains (see Fig.~1).  This quantity satisfies the
initial condition $Q_m(0)=\delta_{m,1}$.  The total domain density,
$N(t)$, is given by $N(t)=\sum_m Q_m(t)$, while the domain
survival probability counts {\em initial} domains that have not shrunk
and hence contains the density $Q_m(t)$ with weight $m$ 
\begin{equation}
\label{stdef}
S(t)=\sum_m m Q_m(t).  
\end{equation}
The average number of domains
contained within a surviving domain $\langle m(t)\rangle=S(t)/N(t)$
grows algebraically according to $\langle m(t)\rangle\sim
t^{\nu-\psi}$ with $\nu$ the domain decay exponent, $N(t)\sim
t^{-\nu}$.  If the behavior of $Q_m(t)$ is truly self-similar, it
should follow the scaling form

\begin{equation}
\label{qmscl}
Q_m(t)\simeq t^{\psi-2\nu}{\cal Q}(mt^{\psi-\nu}).
\end{equation}
The scaling function ${\cal Q}(z)$ exhibits the following extremal
behavior

\begin{equation}
\label{qzscl}
{\cal Q}(z)\sim\cases{z^{\sigma}&$z\ll 1$,\cr 
\exp(-\kappa z)&$z\gg 1$.}
\end{equation}
The small argument tail describes domains that contain a very small
number of initial domains. In particular, the quantity 

\begin{equation}
\label{qtscl}
Q_1(t)\sim t^{-\delta}
\end{equation}
is of special interest: It gives the density of domains which avoided
merging with their neighboring domains up to time $t$.

The  inequalities $Q_1(t)\leq \sum_m Q_m(t)\leq \sum_m m Q_m(t)$
lead to the bounds

\begin{equation} 
\label{bound1}
\psi\leq\nu\leq\delta. 
\end{equation}
Taking into account that at least one surviving domain surrounds a
persistent spin gives $P(t)\leq S(t)$, where $P(t)\sim t^{-\theta}$ 
is the density of  persistent spins.  Thus we arrive at another
upper bound

\begin{equation}
\label{bound}
\psi\leq\theta
\end{equation}
for the exponent $\psi$.  We shall show below that these bounds are
strict for the Potts model and that there are models with
$\psi=\theta$, $\psi=\nu$, and $\nu=\delta$. The bounds (\ref{bound1})
and (\ref{bound}) suggest that a domain decays with the slowest rate
in the problem. 

A useful relation between the scaling
exponents can be obtained by substituting $m=1$ in Eq.~(\ref{qmscl})

\begin{equation}
\label{sr}
\delta-\nu=(\nu-\psi)(1+\sigma).
\end{equation}
Thus, among the three exponents $\psi$, $\delta$, and $\sigma$, only
two are independent. For the $q$-state Glauber-Potts model both the
domain decay exponent $\nu=1/2$\cite{glauber,bbd} and the persistence
exponent $\theta(q)$\cite{dhp} are known. It gives hope that 
analytical determination of the domain exponents is also possible.

Quite obviously, domains disappear when their size vanishes, and
therefore the domain size and number distributions are intimately
related. One is therefore forced to consider the distribution of
domains of size $n$ consisting of $m$ original domains at time $t$,
denoted by $P_{n,m}(t)$. The aforementioned number distribution is
$Q_m(t)=\sum_n P_{n,m}(t)$, and consequently, the domain survival
probability is $S(t)=\sum_{n,m} m P_{n,m}(t)$. 

As will be seen later, studying the joint size-number distribution
requires detailed knowledge of the domain size distribution
$P_n(t)=\sum_m P_{n,m}(t)$. This distribution obeys the normalization
condition 
\begin{equation}
\label{norm}
\sum_n nP_n(t)=1
\end{equation}
reflecting the conservation of the total length. The total domain
density is simply  $N(t)=\sum_n P_n(t)$.  Since the
average domain length grows as $n\sim t^{\nu}$, the length
distribution follows the scaling form
\begin{equation}
\label{pnscl}
P_n(t)\simeq t^{-2\nu}{\cal P}(nt^{-\nu}).
\end{equation} 
The length distribution scaling function has the following limiting
behavior \cite{ab,dz}
\begin{equation}
\label{pxscl}
{\cal P}(x)\sim\cases{x&$x\ll 1$,\cr 
\exp(-\lambda x)&$x\gg 1$.}
\end{equation}
In subsection D, we shall develop an approximation scheme that helps
elucidate many of the qualitative and quantitative features of the
domain size and number distributions.

\subsection{Solvable Cases}

\subsubsection{The $q\to\infty$ Limit}

In the $q=\infty$ case,  similar phase domains never coalesce 
and therefore the domain number is trivial, $m=1$. Thus
$P_{n,m}(t)=P_n(t)\delta_{m,1}$, $N(t)=S(t)=Q_1(t)$, and
$\nu=\psi=\delta=1/2$. The value of the exponent has been obtained by
noting that domain boundaries perform independent random walks and a
domain disappears when its boundaries meet.  Thus domains survive with
probability identical to that of a random walk in the vicinity of a
trap, $N(t)\sim t^{-1/2}$\cite{feller}, or $\nu=1/2$.  On the other
hand, an individual up spin inside this domain has not changed its
sign if it has not been crossed by both domain walls.  Therefore, the
persistence probability is proportional to $t^{-1/2}\times t^{-1/2}
=t^{-1}$, i.e. $\theta=1$. Hence, the bound (\ref{bound}) is strict.

The domain length distribution $P_n(t)$ obeys the diffusion equation
\begin{equation}
\label{difeq}
{dP_n\over dt}=P_{n+1}+P_{n-1}-2P_n,
\end{equation}
with the boundary condition $P_0(t)=0$.  This rate equation satisfies
the length conservation of Eq.~(\ref{norm}).  Solving (\ref{difeq})
subject to the appropriate initial conditions, $P_n(0)=\delta_{n,1}$,
gives

\begin{equation}
\label{exactpnm}
P_n(t)=[I_{n-1}(2t)-I_{n+1}(2t)]\exp(-2t),
\end{equation} 
and 
\begin{equation}
\label{exactN}
N(t)=S(t)=Q_1(t)=[I_0(2t)+I_1(2t)]\exp(-2t),
\end{equation}
where $I_n$ is the modified Bessel function of order $n$\cite{bo}.
The length distribution scales according to Eq.~(\ref{pnscl}), with
$\nu=1/2$ and ${\cal P}(x)\simeq x\exp(-x^2/4)/\sqrt{\pi}$.  The
generic exponential behavior of Eq.~(\ref{pxscl}) is now replaced by a
Gaussian one, indicating that $\lambda\to 0$ as $q\to\infty$.
In the long time limit, $S(t)=Q_1(t)=N(t)\simeq (\pi t)^{-1/2}$
confirming the previous heuristic findings $\nu=\psi=\delta=1/2$.

\subsubsection{The $q\to 1$ Limit}

The 1D $T=0$ Glauber-Potts model with arbitrary $q\geq 1$ can be
mapped to the Ising model with magnetization $\mu=2/q-1$\cite{sm}.  In
other words, the volume fraction of the down phase is $\varphi=1-1/q$
\cite{bfk}.  In particular, the limit $\varphi\to 0$ allows treatment
of the limiting case $q\to 1$ by focusing on the majority domains. The
typical initial size of such domains is $\varphi^{-1}\to\infty$.  This
shows that in the limiting case $q=1$ the minority domains cannot meet
and the majority domains' sizes change appreciably due to coalescence.
Thus, majority domains never disappear, i.e., $S(t)=1$ and $\psi=0$.
Similarly, the persistence exponent is found: $\theta=0$.  A majority
domain remains unreacted till time $t$ if both of its neighboring
minority domains survive, $Q_1(t)=N^2(t)$. The density is given by the
$q=\infty$ solution (\ref{exactN}), and we find $Q_1(t)\simeq (\pi
t)^{-1}$ and $\delta=1$.

The number distribution of the majority domains can be determined as
well.  The dynamics proceeds by minority domains shrinking to zero and
thus leading to coalescence of surrounding majority domains.  Such
aggregation events occur independently with rate $P_1/N^2$ and the
domain number distribution evolves according to

\begin{equation} 
\label{Mnt} 
{dQ_m\over dt}={P_1\over
N^2}\left[\sum_{i=1}^{m} Q_j Q_{m-j} -2NQ_m\right], 
\end{equation}
subject to the initial conditions $Q_m(0)=\delta_{m,1}$.  It is 
helpful to absorb the time-dependent rate $P_1/N^2$ into the time 
variable 

\begin{equation} 
\label{T} 
T=\int_0^t dt'\,{P_1(t')\over N^2(t')}=N^{-1}(t)-1, 
\end{equation}
with the overall density of (\ref{exactN}), and the last equality
evaluated using $\dot N=-P_1$.  With this time variable 
Eq.~(\ref{Mnt}) reduces to the classical Smoluchowski equation
\cite{smol}

\begin{equation} 
\label{CMnt}
{dQ_m\over dT}=\sum_{j=1}^{m} Q_j Q_{m-j}-2NQ_m.  
\end{equation}
Solving Eq.~(\ref{CMnt}) with the appropriate monodisperse initial
conditions gives ${\cal Q}_m(T)=T^{m-1}(1+T)^{-m-1} \simeq
T^{-2}\exp(-m/T)$\cite{smol}.  Indeed $Q_1(T)=N^2(T)=(1+T)^{-2}$, in
agreement with the previous argument.  In the long-time limit,
$T\simeq N^{-1}\simeq \sqrt{\pi t}$ and $Q_m(t)\simeq (\pi
t)^{-1}\exp\left[-m(\pi t)^{-1/2}\right]$.  Thus the domain number
distribution scales according to Eq.~(\ref{qmscl}) with the purely
exponential scaling function

\begin{equation} 
{\cal Q}(z)=\pi^{-1}\exp(-z\pi^{-1/2}).  
\end{equation}  
The average domain properties are $S(t)=1$, $N(t)\simeq (\pi
t)^{-1/2}$, and $Q_1(t)=N^2(t)\sim (\pi t)^{-1}$ and the
scaling exponents $\sigma=\psi=0$, $\nu=1/2$, $\delta=1$.

Changes in the domain size due to domain wall diffusion are negligible
here and the joint size-number distribution evolves according to

\begin{equation} 
\label{pnmT}
{dP_{n,m}\over dT}=
\sum_{i,j}P_{i,j}P_{n-i,m-j}-2NP_{n,m}.
\end{equation}
Eq.~(\ref{pnmT}) generalizes the Smoluchowski equations for
aggregation with two conservation laws\cite{aggr}.  Introducing the
generating function $F(u,v,T)=\sum_{n,m} u^n v^m P_{n,m}(T)$ one can
solve Eq.~(\ref{pnmT}) for {\em arbitrary} initial conditions to
find\cite{aggr} $F(u,v,T)=F_0(u,v)(1+T)^{-1}[1+T-TF_0(u,v)]^{-1}$.  In
the present case, the appropriate initial conditions are
$P_{n,m}(0)=\delta_{m,1}\varphi^2(1-\varphi)^{n-1}$ and hence
$F_0(u,v)=uv\varphi^2[1-u(1-\varphi)]^{-1}$.  Evaluating the 
limits $n\varphi\to n$, and $t\to\infty$, we arrive at the
scaling form $P_{n,m}(t)\simeq t^{-5/4}\Phi(x,y)$, with the scaling
variables $x=(m+n)(\pi t)^{-1/2}$, $y=(m-n)(\pi t)^{-1/4}$, and the
scaling function
\begin{equation} 
\label{pT}
\Phi(x,y)=(\pi x)^{-1/2}\exp(-x-y^2/2x).  
\end{equation}
Instead of the naive scaling variables $n t^{-1/2}$ and $m t^{-1/2}$,
unusual scaling variables underlie the scaling function (\ref{pT}).
The former scaling variable $x$ is just the sum of the naive scaling
variables while the latter ``diffusive'' scale $y$ is hidden. In this
case, the domain length and the domain number are equivalent, and
their underlying scaling functions are identical ${\cal
  P}(x)=\pi^{-1}\exp(-x\pi^{-1/2})$.  Nevertheless, there is a
considerable difference between the variables $n$ and $m$ as the
latter is generally not a conserved quantity.  Furthermore, as $q$
decreases from $\infty$ to $1$, the decay coefficient $\lambda$
governing the domain length distribution increases from $0$ to
$\pi^{-1/2}$.

The above results apply for $q$ close to unity as long as neighboring
domains do not interact, i.e., as long as the diffusion time scale is
smaller that the domain size, $\sqrt{t}\ll\varphi^{-1}$.  Eventually,
this no longer holds, and correlations between majority domains
develop. Nevertheless, in the limit $q\to 1$ the equations
(\ref{CMnt}) and (\ref{pnmT}) are {\em exact} since no correlations
develop if none are present initially.  Similar reasoning applies to
several models where domains are immobile and merging occurs
\cite{dbg1}.

\subsection{Independent Interval Approximation (IIA)}

Ignoring correlations between neighboring domains allows us to develop
an approximate theory for the time-evolution of the domain
distribution. This so-called IIA proved useful in studies of related
reaction-diffusion processes\cite{ab,msbc,bbd}.

\subsubsection{The Length Distribution}

The joint number distribution requires knowledge of the length
distribution and we start by deriving a master equation for $P_n(t)$.
Consider first the Ising case.  In an infinitesimal time interval
$\Delta t$, the domain  $P_n(t)$ changes according to
\begin{eqnarray} 
\label{differ} 
P_n(t&+&\Delta t)=(1-2\Delta t)P_n(t)\nonumber\\ 
&+&\Delta t\,P_{n-1}(t)\left[1-{P_1(t)\over N(t)}\right] 
+\Delta t\,P_{n+1}(t)\\ &-&\Delta t\,P_n(t)\,{P_1(t)\over N(t)} 
+\Delta t\,P_1(t)
\sum_{i+j+1=n}{P_i(t)\over N(t)}{P_j(t)\over N(t)}\nonumber, 
\end{eqnarray} 
where $N(t)=\sum_{n}P_n(t)$ is the total domain density.  The first
term on the right-hand side of Eq.~(\ref{differ}) counts for the
probability that both domain walls do not hop.  The next two terms
describe gain due to diffusion, with the prefactor $(1-P_1/N)$ in the
second term to ensure that the hopping domain wall does not disappear.
The forth term represents loss due to disappearance of the smallest
domain, located on the boundary of our domain, while the final term
accounts for gain due to domain merger. 

Eq.~(\ref{differ}) assumes that the sizes of adjacent domains are
uncorrelated, and thus is mean-field in nature.  In the limit $\Delta
t\to 0$ the difference equations (\ref{differ}) turn into a system of
differential equations:

\begin{eqnarray}
\label{ising}
{dP_n\over dt}&=&P_{n-1}+P_{n+1}-2P_n\\
&+&{P_1\over N^2}\left[\sum_{i=1}^{n-2} P_i P_{n-1-i}-
N(P_n+P_{n-1})\right].\nonumber
\end{eqnarray}
Eqs.~(\ref{ising}) apply for $n=1$ if we set $P_0\equiv 0$.  One
crucial test is to verify the length conservation of Eq.~(\ref{norm}).
Another test is to sum all equations in (\ref{ising}) to get $\dot
N=-2P_1$.  This is an {\em exact} equation, since three domains
disappear and one is born in each annihilation event.

Generally, in the Potts case, the domain size distribution evolves
according to the rate equation

\begin{eqnarray}
\label{pn}
{dP_n\over dt}=&&P_{n-1}+P_{n+1}-2P_n\\
+&&{P_1\over (q-1)N^2}\left[\sum_{i=1}^{n-2} P_i P_{n-1-i}
-N(P_n+P_{n-1})\right].\nonumber
\end{eqnarray}
Indeed, collision of domain walls results in annihilation with
probability ${1\over q-1}$ or in coalescence with probability ${q-2
  \over q-1}$. Only annihilation events affect the domain distribution
and thus the ${1\over q-1}$ prefactor of the annihilation term.  In
the cases $q=2$ and $q=\infty$, Eqs.~(\ref{ising}),(\ref{difeq}) are
clearly recovered.  In the limit $q\to 1$, only the reaction term
survives, in agreement with Eq.~(\ref{pnmT}).  One can also verify
that the total length is conserved and the total domain density decays
according to the {\it exact} rate equation 

\begin{equation}
\label{N}
{dN\over dt}=-{q\over q-1}P_1.
\end{equation}

The diffusion term in Eq.~(\ref{pn}) implies $\langle n(t)\rangle\sim
t^{1/2}$, and since $\langle n\rangle\sim N^{-1}$ the correct decay
exponent $\nu=1/2$ \cite{glauber,bbd} is recovered.  In the following,
we will need to determine the asymptotic prefactor $A$,  $N(t)
\sim At^{-1/2}$, $A=\int dx\,{\cal P}(x)$.  The density rate equation
(\ref{N}) implies $P_1\simeq {\cal P}'(0)t^{-3/2}$ with ${\cal
P}'(0)={q-1\over 2q}A$.

A quantitative analysis of Eq.~(\ref{pn}) may be carried by treating
the variable $n$ as continuous. The quantity ${\cal P}(x)$ satisfies

\begin{equation}
\label{px}
{\cal P}''+{1\over 2}(x{\cal P})'+{q-2\over 2q}{\cal P}
+{1\over 2qA}\,{\cal P}*{\cal P}=0,
\end{equation}
where ${\cal P}'\equiv d{\cal P}/dx$ and ${\cal P}*{\cal
P}\equiv\int_0^x dy\,{\cal P}(y){\cal P}(x-y)$. The 
normalized Laplace transform of the scaling function ${\cal P}(x)$, 
$p(s)=A^{-1}\int_0^{\infty} dx\,e^{-sx}\,{\cal P}(x)$, obeys 

\begin{equation}
\label{riccati}
{dp\over ds}={p^2\over qs}+\left(2s+{q-2\over qs}\right)p
-{q-1\over qs},
\end{equation}
subject to the boundary condition $p(0)=1$.  The transformation
$p(s)=1-qs^2-qs{d\over ds}\ln y(s)$ reduces the Riccati equation
(\ref{riccati}) into the parabolic cylinder equation,

\begin{equation}
\label{par}
{d^2 y\over ds^2}+\left(1+{2\over q}-s^2\right)y=0. 
\end{equation}
The solution to (\ref{par}) reads
$y(s)=C_-D_{1/q}(-s\sqrt{2})+C_+D_{1/q}(s\sqrt{2})$, with 
$D_{1/q}(x)$ the parabolic cylinder function of order $1/q$ \cite{bo}.
The large $s$ behavior of $p(s)$, $p(s)\sim {q-1\over 2q}s^{-2}$,
implies $C_-=0$, and we get

\begin{equation}
\label{ps}
p(s)=1-qs^2-qs{d\over ds}\ln D_{1/q}(s\sqrt{2}).
\end{equation}
The normalization condition $\sum_n nP_n(t)=1$ can be
reduced to $Ap'(0)=-1$.  This allows us to determine the constant

\begin{equation}
\label{A}
A={\Gamma\big[1-{1\over 2q}]
\over\Gamma\big[{1\over 2}-{1\over 2q}\big]},
\end{equation}
where $\Gamma$ denotes the gamma function.  In deriving (\ref{A}) 
we have used the properties\cite{bo}

\begin{equation}
\label{large}
D_c(x)\sim x^c\exp(-x^2/4)[1+{\cal O}(x^{-2})],
\end{equation}
and
\begin{equation}
D_c(0)={\pi^{1/2}2^{c/2}\over \Gamma(1/2-c/2)},\quad
D_c'(0)=-{\pi^{1/2}2^{(c+1)/2}\over \Gamma(-c/2)}.
\end{equation}

The value of the prefactor $A$ predicted by the IIA may be compared to
the exact one, $A_{\rm exact} =(1-q^{-1})/\sqrt{\pi}$ \cite{amar} (see
Fig.2).  In the extreme cases of $q=1$ and $q=\infty$ the prefactor $A$
is exact.  The mismatch is worst for the Ising ($q=2$) case where
$A=\Gamma(3/4)/\Gamma(1/4)\cong 0.337989$ while $A_{\rm
exact}=(4\pi)^{-1/2}\cong 0.28209$\cite{glauber}.

\begin{figure}
\vspace{-.25in}
\centerline{\epsfxsize=7cm \epsfbox{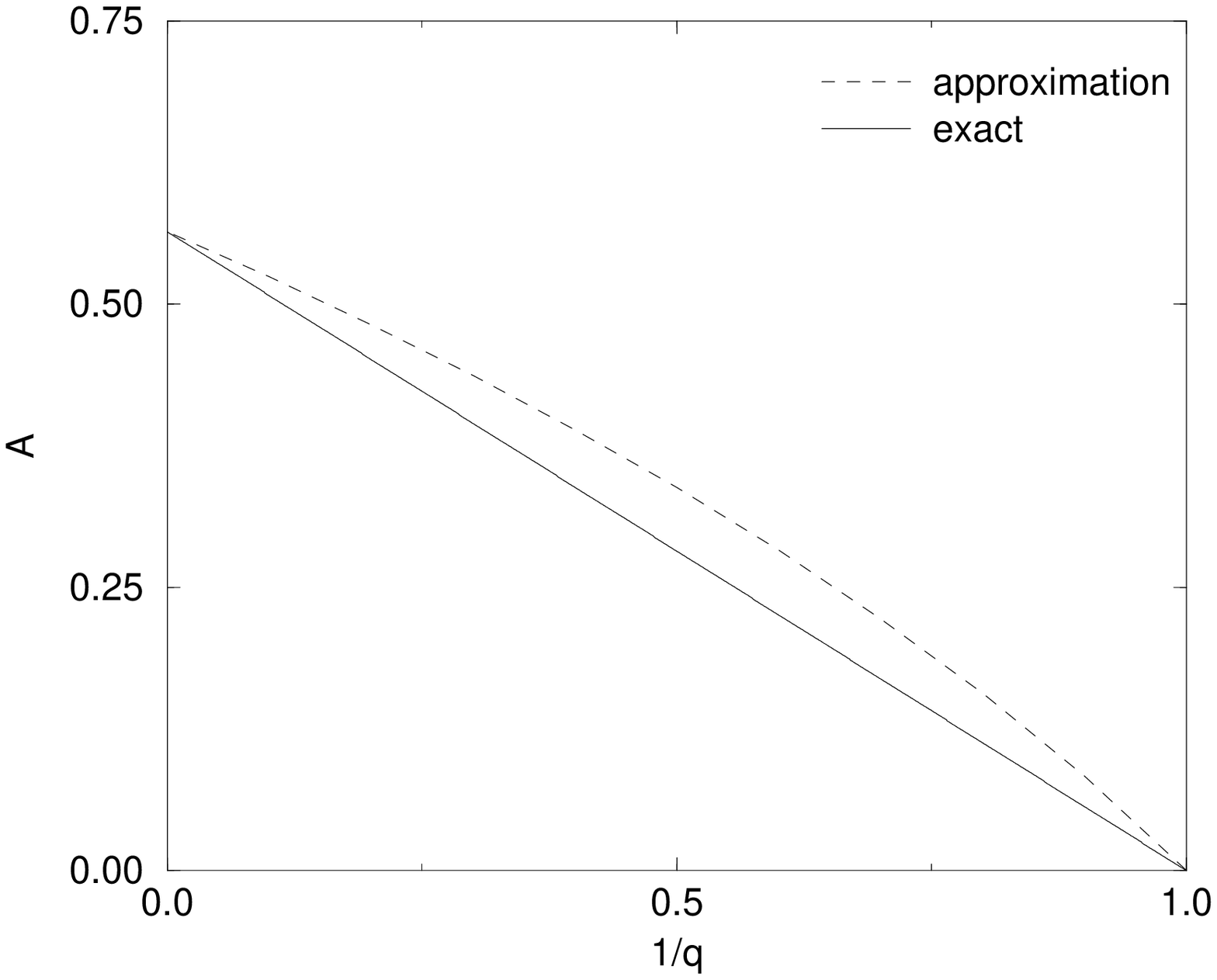}}
\noindent
{\small {\bf Fig.~2}.  The prefactor $A$ of Eq.~(\ref{A}) vs. the exact
value $A_{\rm exact}=(1-q^{-1})/\sqrt{\pi}$.}
\end{figure}

The IIA gives correct qualitative results including: (i) density decay
$N(t)\sim t^{-1/2}$; (ii) linear small size distribution, ${\cal
  P}(x)\simeq {A(q-1)\over 2q}x$, as can be seen by considering the
large $s$ behavior of $p(s)\simeq {(q-1)\over 2q}s^{-2}$; (iii)
exponential tail, ${\cal P}(x)\simeq qA\lambda\exp(-\lambda x)$ as in
Eq.~(\ref{pxscl}).  The tail follows from the behavior of the Laplace
transform $p(s)\simeq q\lambda/(s+\lambda)$ near its pole at negative
$s=-\lambda$, given by the first zero of
$D_{1/q}(-\lambda\sqrt{2})=0$.  For the Ising case one has
$\lambda=0.5409$. This value should be compared with the exact value
$\lambda=\zeta(3/2)/4\sqrt{\pi} =0.368468$ obtained by Derrida and
Zeitak\cite{dz} and the approximate value $\lambda=0.35783$ obtained
by Alemany and ben-Avraham \cite{ab}.

One may wonder regarding the value of the IIA.  It is analogous to
that of Ref.~\cite{ab} but the quantitative agreement is worse in our
case.  The answer is simple: Our approach is {\em self-consistent}
while the approach of Ref.~\cite{ab} is not.  Indeed, making use of
the assumption that the domain sizes are uncorrelated, Alemany and
ben-Avraham express the equal-time two-spin correlation function via
the domain size distribution $P_n(t)$.  Then they use the exact
expression for the equal-time two-spin correlation function
\cite{glauber} to determine $P_n(t)$.  However, would they use their
key assumption that the domain sizes are uncorrelated everywhere they
would eventually obtain our expression for $P_n(t)$.  In contrast, our
approach is self-consistent as all our the results are derived within
the same scheme.  Additionally, to determine the more subtle 
characteristics to be described below one does not enjoy the luxury of
known exact analytical results.

\subsubsection{The Joint Distribution}

We are now in a position to tackle the joint size-number distribution,
$P_{n,m}(t)$, which captures both the spatial and ``historical''
characteristics of the coarsening domain mosaic.  The corresponding
rate equation is a generalization of Eq.~(\ref{pn})

\begin{eqnarray} 
\label{pnm}
&&{dP_{n,m}\over dt}=P_{n-1,m}+P_{n+1,m}-2P_{n,m}\\
&&+{P_1\over (q-1)N^{2}}\left[\sum_{i,j}P_{i,j}P_{n-1-i,m-j}
-N(P_{n,m}+P_{n-1,m})\right] \nonumber
\end{eqnarray}
with the initial condition $P_{n,m}(0)=\delta_{n,1}\delta_{m,1}$ and
the boundary condition $P_{0,m}(t)=0$.  The variable $m$ is ``mute''
in some sense. It appears in a nontrivial way only in the convolution
term.  One should verify that this master equation is self-consistent.
First, by summing over $m$, we recover Eq.~(\ref{pn}).  Second, it
implies that the domain survival probability satisfies the exact
linear equation ${dS/dt}=-\sum_m m P_{1,m}$.  So far, we have not
succeeded in solving the joint distribution. 
Nevertheless, it is still possible to obtain analytically many
interesting properties of Eq.~(\ref{pnm}), including the scaling
exponents.

Let us consider the distribution of domains which have not merged with
other domains up to $t$, $R_n(t)\equiv P_{n,1}(t)$. For such domains,
the convolution term vanishes and they evolve according to the linear
rate equation 
\begin{equation}
\label{rn}
{dR_n\over dt}=R_{n-1}+R_{n+1}-2R_n-{P_1\over (q-1)N}(R_n+R_{n-1})
\end{equation}
with the initial condition $R_n(0)=\delta_{n,1}$ and the boundary
condition $R_0(t)=0$.  In the continuum limit we again replace
$R_{n-1}+R_{n+1}-2R_n$ by $\partial^2 R/\partial n^2$ and
$R_n+R_{n-1}$ by $2R_n$ to find a diffusion-convection equation for
$R_n(t)$.  The transformation $R_n\to \tilde R_n N^{-2/q}$ reduces
this equation to the diffusion equation (\ref{difeq}) for $\tilde
R_n$, which is solved to yield $R_n(t)\simeq N^{2/q}t^{-1}{\cal
  R}(nt^{-1/2})$, with ${\cal R}(x)=x\exp(-x^2/4)/\sqrt{\pi}$.  The
large $n$ behavior of $P_{n,1}(t)$ mimics the $q=\infty$ case in that
it exhibits a Gaussian behavior, $P_{n,1}(t)\sim \exp(-n^2/4t)$, while
the average domain density decays exponentially, $P_n(t)\sim
\exp(-n/\sqrt{t})$.  Large intervals are more strongly suppressed when
they have a small number of ancestors and therefor, the $n$-tail of
the joint distribution strongly depends upon $m$.  The total density
of unreacted domains is $Q_1(t)=\sum_n R_n\sim t^{-{1\over 2}-{1\over
    q}}$, which gives the decay exponent
\begin{equation}
\label{delta}
\delta={1\over 2}+{1\over q}.  
\end{equation}

Obtaining the second independent exponent $\psi$ is more involved. The
natural approach, i.e., a direct investigation of the domain number
distribution, $Q_m$, appears to be useless, as it requires knowledge
of $P_{1,m}$ and hence the entire $P_{n,m}$. The domain survival
probability can be alternatively obtained by considering
$U_n(t)=\sum_m mP_{n,m}(t)$. This quantity obeys

\begin{eqnarray}
\label{un}
{dU_n\over dt}&=&U_{n-1}+U_{n+1}-2U_n\\
&+&{P_1\over (q-1)N^2}
\left[2\sum_{i=1}^{n-2} U_i P_{n-1-i}-N(U_n+U_{n-1})\right],\nonumber
\end{eqnarray}
obtained by summing Eqs.~(\ref{pnm}).  We write $U_n(t)$ in a scaling
form $U_n(t)\simeq t^{-\psi-1/2}{\cal U}(nt^{-1/2})$.  Asymptotically,
the domain survival probability reads $S(t)\simeq Bt^{-\psi}$ with
$B=\int dx\, {\cal U}(x)$.  The scaling distribution satisfies

\begin{equation}
\label{ux}
{\cal U}''+{1\over 2}(x{\cal U})'+ \left(\psi-{1\over
q}\right){\cal U} +{1\over qA}{\cal U}*{\cal P}=0.  
\end{equation}
The  normalized Laplace transform of the scaling function ${\cal U}(x)$,
$u(s)=B^{-1}\int_0^\infty dx\,e^{-sx}{\cal U}(x)$, obeys

\begin{equation}
\label{dus}
{du\over ds}=2\left({p(s)+q\psi-1\over qs}+s\right)u
-{2\psi\over s}, \quad u(0)=1.
\end{equation}
In deriving (\ref{dus}) we used the relation ${\cal U}'(0)=B\psi$,
found by integration of Eq.~(\ref{ux}), combined with
$A=\int dx {\cal P}(x)$. Substituting the explicit expression
(\ref{ps}) for $p(s)$ into Eq.~(\ref{dus}), and solving for $u(s)$
yields
\begin{equation} 
\label{us} u(s)=2\psi
s^{2\psi}D^{-2}_{1/q}(s\sqrt{2}) \int_s^{\infty}\!\!dr\,
r^{-2\psi-1}D_{1/q}^{2}(r\sqrt{2}).  
\end{equation} 
This solution is consistent with the anticipated $s\to\infty$
behavior, $u(s)\simeq \psi s^{-2}$. Furthermore, evaluating
Eq.~(\ref{us}) near the origin gives
$u(s)=1+F(\psi)s^{2\psi}+Cs+\cdots$.  Therefore for $u'(s)$ to be
finite near $s=0$, we must have $F(\psi)=0$. Evaluating $F(\psi)$
gives

\begin{equation} 
\label{psi} 
0=\int_0^\infty dr\,r^{-2\psi}D_{1/q}(r)D_{1/q}'(r),
\end{equation} 
Interestingly the second domain survival exponent, $\psi$, is 
irrational, in contrast with $\delta$.

For completeness, we write the leading extremal behavior of the
function $U(x)$:
 
\begin{equation} 
\label{uxext} 
U(x)\sim\cases{x&$x\to 0$;\cr 
x\exp(-\lambda x)&$x\to \infty$.\cr}
\end{equation}
This behavior can be easily obtained from the extremal
behavior of the function $u(s)$. When $s\to\infty$, $u(s)\sim s^{-2}$,
while near the pole $s\to -\lambda$, one finds $u(s)\sim
(s+\lambda)^{-2}$.

\subsection{Numerical Results}

To test the IIA predictions, we performed numerical simulations on a
spin chain of size $L=10^7$.  Random initial conditions and periodic
boundary conditions were used.  The simulation data represents an
average over 10 different realizations. For the Ising case, we found
the exponent values $\psi=0.126(1)$ and $\delta=1.27(2)$ (see Figs. 3
and 4).  These values should be compared with the IIA predictions of
$\psi=0.136612$ and $\delta=1$.

As was the case for the persistence exponent, $\theta$, the domain
exponents strongly depend on $q$.  Numerical values of the exponents
$\psi$ and $\delta$ are summarized in Table 1 for representative
values of $q$.  As $q$ increases, the approximation improves and
eventually becomes exact for the extreme case $q=\infty$. Thus,
$\psi$ is overestimated by up to 10\% and $\delta$ is underestimated
by up to 25\%. Hence, domains of average number $m$ are better
approximated in comparison with domains with extremely small $m$.
Although the estimates are not exact they are still useful as they
exhibit the correct $q$-dependence.  In the $q\to\infty$ limit, the
exponents approach the limiting value 1/2 according to $\delta={1\over
  2}+{1\over q}$ derived in Eq.~(\ref{delta}), and $\psi\cong {1\over
  2}-{1\over q}$.  The leading behavior for $\psi$ follows directly
from the scaling relation (\ref{sr}), $\delta$ and $\sigma(\infty)=0$.

In the $q\to 1$ limit analysis of Eq.~(\ref{psi}) suggests that $\psi$
vanishes according to $\psi\propto (q-1)^2$. This behavior agrees with
the $q=1$ exact solution and is consistent with simulation results of
an Ising chain with magnetization $\mu=2/q-1$. It is practically
impossible to obtain $\delta$ conclusively because unreacted domains
decay quickly.  The limiting value as $q\to 1$ appears to be larger
than the value suggested by the IIA, $\delta\cong 3/2-(q-1)$.

We performed several checks to verify that the asymptotic behavior of
Eq.~(\ref{st}) and Eq.~(\ref{qtscl}) is robust.  For example, it is
independent of the initial domain wall concentration (provided that
the correlations in the initial condition are short range). We
conclude that $\psi$ and $\delta$ are nontrivial exponent, i.e., they
cannot be extracted from the known exponents associated with the
Ising-Glauber model.  Similar to the persistence exponent,
$\theta(q)$, the exponents appear to be irrational except for the
limiting cases $q=\infty$ ($\psi=\delta={1\over 2}$, $\sigma=0$) and,
maybe, for $q=2$ ($\psi={1\over 8}$, $\delta={5\over 4}$, $\sigma=1$).

The numerical simulations also confirm that the distribution function
$Q_m(t)$ scales according to Eq.~(\ref {qmscl}) (see Fig. 5). The
scaling function ${\cal Q}(z)$, defined in Eq.~(\ref{qzscl}), decays
exponentially for large argument and is algebraic for small
argument. The scaling relations combined with the simulation values
give $\sigma=1.05(5)$. This is consistent with the linear behavior
seen in Fig.~5 for $z\ll 1$.  Hence, similar scaling functions
underlie the domain number and size distributions \cite{ab,dz}. As $q$
increases from 2 to $\infty$, the exponent $\sigma$ decreases from $1$
to $0$, respectively.

\begin{figure}
\vspace{-.30in}
\centerline{\epsfxsize=9cm \epsfbox{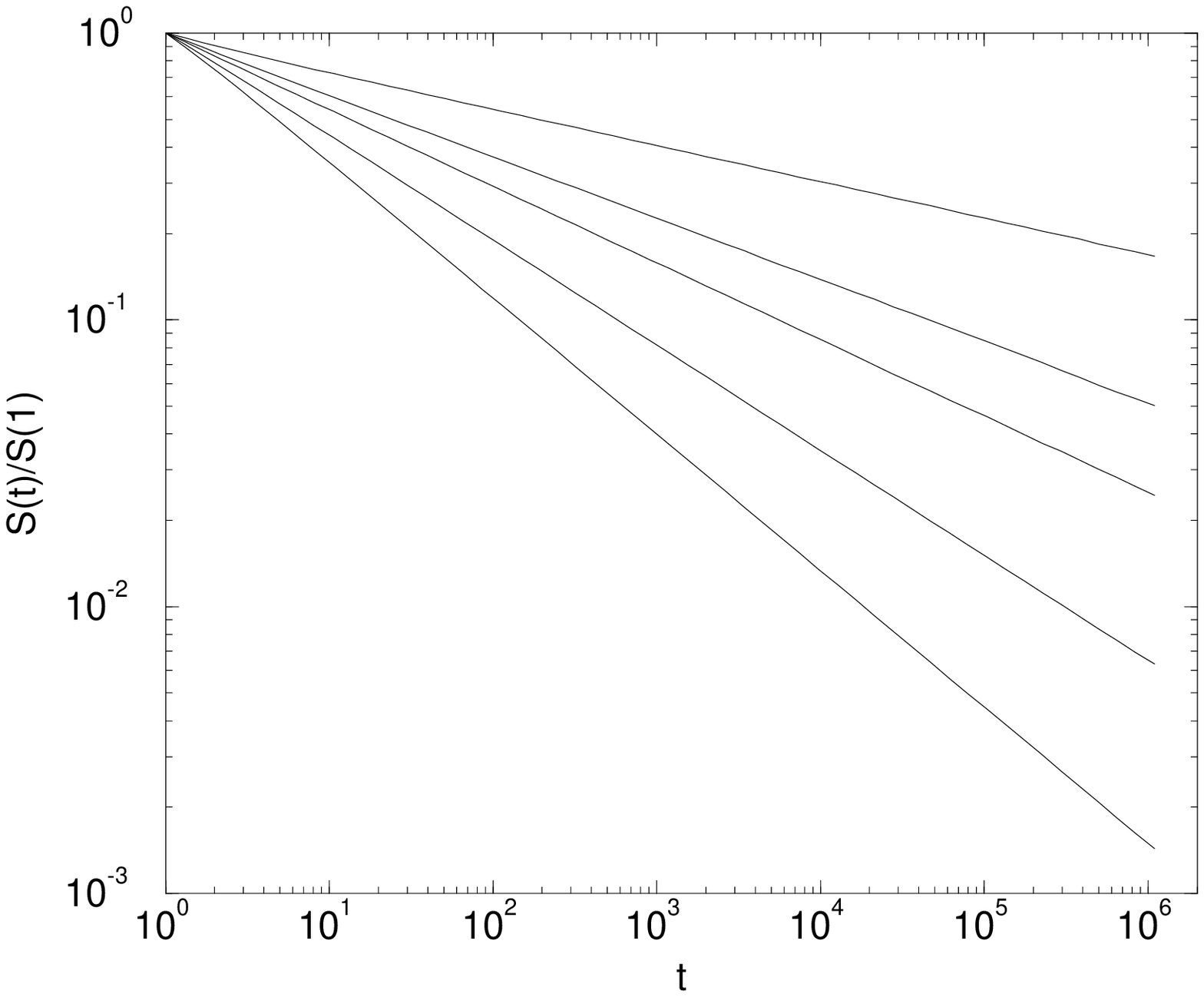}}
\vspace{-.25in} 
\noindent
{\small {\bf Fig.~3}. The domain survival probability $S(t)/S(t=1)$ in
the Potts model. Shown are Monte Carlo simulation results for
$q=2,3,4,8$ and 50.}
\end{figure}

\begin{figure}
\vspace{-.30in}
\centerline{\epsfxsize=9cm \epsfbox{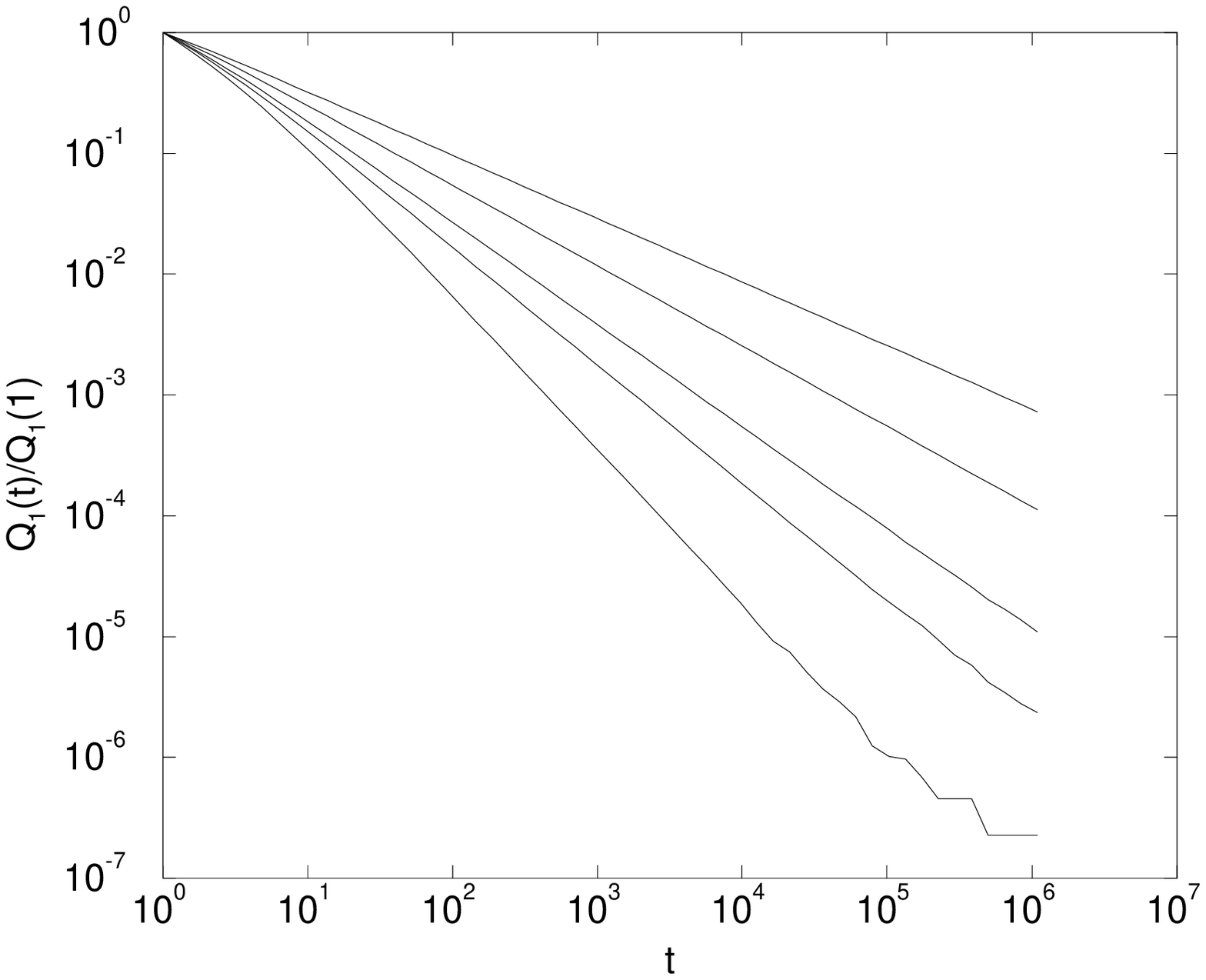}}
\vspace{-.25in} 
\noindent
{\small {\bf Fig.~4}. The density of unreacted
domains $Q_1(t)/Q_1(t=1)$ in the Potts model.  The same 
Monte Carlo simulation as in Fig.~3.}
\end{figure}

\vspace{.1in} 
\centerline{
\begin{tabular}{|l|c|c|c|c|c|}
\hline
&\multicolumn{3}{c|}{MC}&\multicolumn{2}{c|}{Eq.~(\ref{pnm})}\\
\hline
$q$&$\psi$&$\delta$&$\sigma$&$\psi$&
$\delta$\\
\hline
2        &\,0.126\,&\,1.27\,&\,1.05\,&\,0.136612\,&\,1\,\\
3        &\,0.213\,&\,0.98\,&\,0.67\,&\,0.231139\,&\,5/6\,\\
4        &\,0.267\,&\,0.85\,&\,0.50\,&\,0.287602\,&\,3/4\,\\
8        &\,0.367\,&\,0.665\,&\,0.24\,&\,0.385019\,&\,5/8\,\\
50       &\,0.476\,&\,0.525\,&\,0.03\,&0.480274 &13/25\\
$\infty$   &\,  1/2\,&\,1/2 \,&\,0\,&\,  1/2\,&1/2\\
\hline
\end{tabular}}
\vspace{.1in} 
\noindent{\small {\bf Table 1}: Domain exponents for the $q$-state
  Potts model in one dimension. Local slopes analysis was applied to
  the simulation data.  The theoretical
  $\psi$ is from Eq.~(\ref{psi}) and $\delta={1\over 2}+{1\over q}$.}

\begin{figure}
\vspace{-.30in}
\centerline{\epsfxsize=9cm \epsfbox{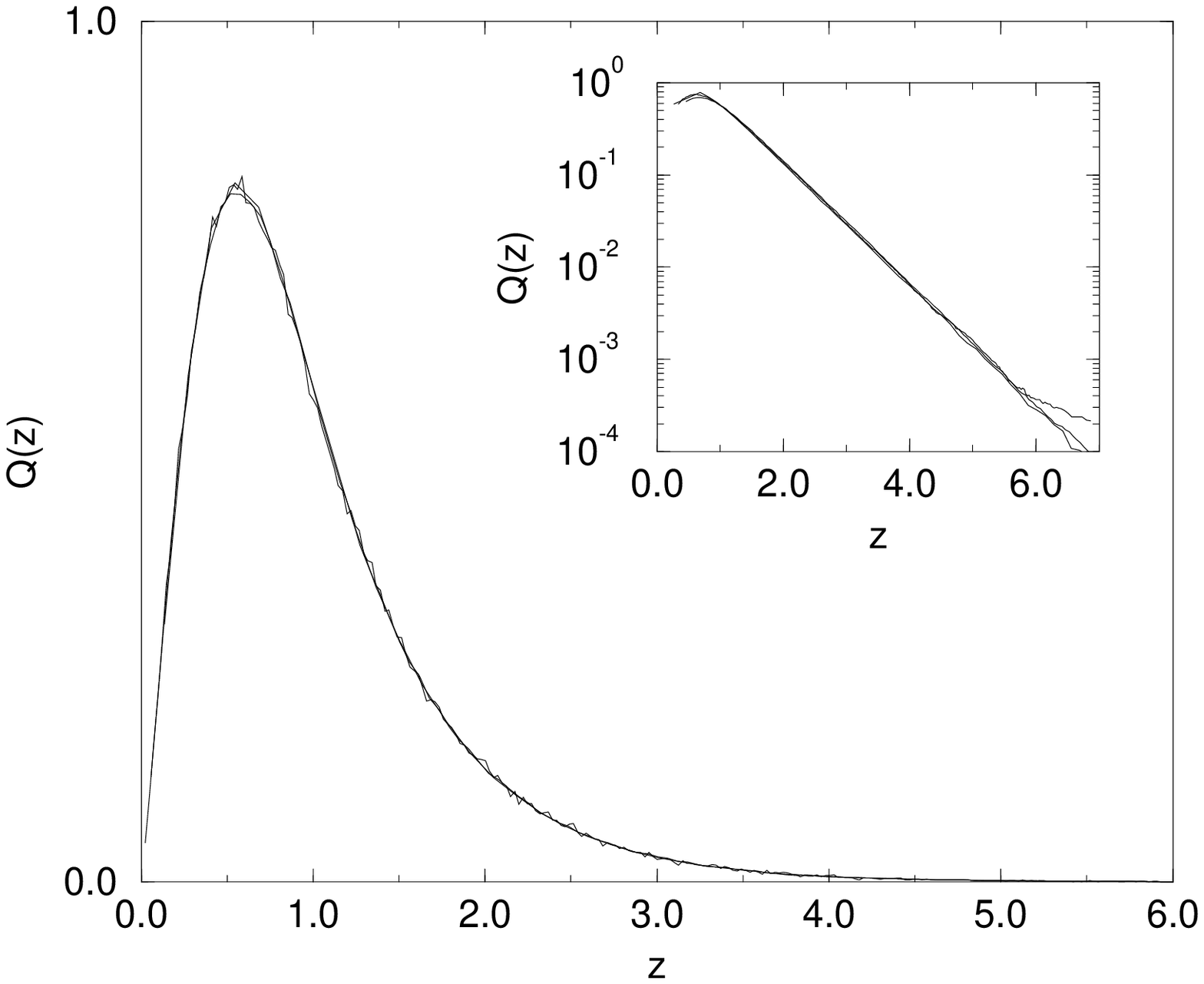}}
\vspace{-.25in} 
\noindent
{\small {\bf Fig.~5}. The scaling distribution ${\cal Q}(z)$ vs. $z$ for
three different times $t=10^2,10^3,10^4$.  The data represents an
average over 100 systems of size $L=10^5$. The inset demonstrates the
exponential behavior of the large-$z$ tail.}
\end{figure}

Direct numerical integration of Eq.~(\ref{pnm}) reveals a number
distribution, $Q_m(t)$, that scales according to Eq.~(\ref{qmscl}),
and has an exponential tail in agreement with the simulation
results. Moreover, the emerging $S(t)$ falls within $5\%$ of the
actual survival probability over a significant temporal range, 
$t<10^3$.  In summary, in addition to predicting
the correct scaling behavior, Eq.~(\ref{pnm}) provides a good
approximation for many quantitative features of the domain
distribution, and in particular, good estimates for the decay
exponents.

\subsection{$q=\infty$ Potts Model in Higher Dimensions}

So far, our discussion has been restricted to one dimension. Domains
are not necessarily well-defined in higher dimensions. In particular,
in the Ising case it is not clear whether our results can be properly
generalized to higher dimensions.  However, in the $q\to\infty$ limit
of the Potts model domains are well defined and dynamics considerably
simplifies\cite{sm}.  In this limit, it has been argued heuristically
and confirmed numerically\cite{dds,ld} and experimentally \cite{tam}
for evolving soap froth, that $\theta=1$ in two dimensions.

We now present a simple heuristic argument which gives the exponents
$\psi=\delta=\theta=d/2$ for the $q=\infty$ Potts model in $d\geq 2$
dimensions.  First we note that an exact correspondence between the
dynamics of the $q$-state Potts model and the Ising model with
magnetization $\mu=2/q-1$ holds only in one dimension.  This happens
due to the {\em global} conservation of the
magnetization\cite{glauber}, a peculiar property of the 1D Ising model
with zero-temperature Glauber dynamics.  This global conservation for
the locally non-conserved dynamics does not happen when $d>1$, as seen
by considering a single up spin in the sea of down spins.  On the
other hand, for the Potts model with symmetric initial conditions the
density of any phase is globally conserved.  This suggests a
correspondence between the Potts model and the Ising model with
globally conserved dynamics.  It appears difficult to make such
correspondence rigorous, although it is supported by several
tests\cite{sm}.  The reduction to the Ising model with magnetization
$\mu=2/q-1$ in $d$ dimensions can be hardly considered as a
simplification except for the $q=\infty$ case which may be analyzed
within the framework of the Lifshitz-Slyozov theory\cite{bray}.
Indeed, as $q\to\infty$ the minority (up) phase approaches
infinitesimal concentration, so up domains do not interact and the
Lifshitz-Slyozov approach, suitably modified for the present
case\cite{sm}, should be exact.

So consider a set of bubbles (the domains tend to become round in high
dimensions) of up phase in a sea of down phase.  We cannot restrict
ourselves to the single-domain situation as in one dimension since
when $d>1$ a single bubble would not evolve, while the set of bubbles
do evolve: small bubbles shrink and large bubbles grow.  We are not
interested in details of the bubble evolution; the only relevant
feature is that the radii distribution scales asymptotically, $N(R,t)$
$N(R,t)\simeq {\cal R}^{-(d+1)}{\cal N}(R/{\cal R})$, with the average
radius ${\cal R}\sim \sqrt{t}$. This behavior is due to the nature of
non-conserved dynamics\cite{bray} and the prefactor guarantees a
conserved magnetization.  Clearly, $S(t)$ is determined by computing
the number of surviving bubbles, $S(t)\sim \int dR N(R,t)\sim {\cal
  R}^{-d}$, implying
\begin{equation}
\label{psid}
\psi=\delta=\theta=d/2.  
\end{equation}

\section{Exactly Solvable Cases} 

Given that obtaining the exact domain distribution in the $q$-state
Potts model appears to be a difficult problem, it might prove useful
to study simpler problems which are exactly solvable. We present in
this section exact results for a variant of the Potts model with
simplified dynamics and for ballistic annihilation.

\subsection{Diffusionless Dynamics}

Consider the $T=0$ $q$-state Potts with simplified dynamics where only
energy lowering transitions are allowed. Thus, domain wall diffusion 
$A0\rightleftharpoons 0A$ in Eq.~(\ref{aa}) is forbidden and the
reaction scheme is 
\begin{eqnarray}
\label{aas}
AA&\buildrel {1\over q-1} \over \longrightarrow &00,\\
AA&\buildrel {q-2\over q-1}\over \longrightarrow&A0\ {\rm or }\ 0A. 
\nonumber
\end{eqnarray}
When $q=2$, exchanging the roles of domain walls and vacant sites,
this problem is equivalent to random sequential adsorption of dimers.
Similarly, the $q=\infty$ case reduces to monomer adsorption subject
to a volume exclusion constrain\cite{jwe}.

Assuming that neighboring intervals are uncorrelated, the domain 
length density rate equation reads
\begin{eqnarray}
\label{pns}
{dP_n\over dt} &=& {P_1\over (q-1)N^2}
\left[\sum_{i=1}^{n-2} P_iP_{n-1-i}-2NP_n\right]\nonumber\\
&&+{(q-2)P_1\over (q-1)N}\left[P_{n-1}-P_n\right]-\delta_{n,1}P_1.
\end{eqnarray}
For simplicity, we consider the antiferromagnetic initial condition
$P_n(0)=\delta_{n,1}$.  While the annihilation term is similar to
Eq.~(\ref{pn}), coalescence events are no longer offset by domain wall
diffusion and thus the second term which is proportional to the
coalescence probability $(q-2)/(q-1)$. It can verified that the total
length $\sum_n n P_n=1$ is conserved.

The density decays according to the familiar rate equation (\ref{N})
$\dot N=-{q\over q-1}P_1$. On the other hand, the minimal gap density
satisfies $\dot P_1=-P_1\left[1+{qP_1\over (q-1)N}\right]$.  It is
useful to introduce the normalized quantity $p_1=P_1/N$ which obeys
$\dot p_1=-p_1$ and thus $p_1=e^{-t}$.  Using this result, the minimal
gap density and hence the density is found

\begin{equation}
\label{nts}
N(t)=\exp[-q(1-e^{-t})/(q-1)]. 
\end{equation} 
This agrees with known exact results for the $q=2$ and $q=\infty$
cases\cite{kv,mp,bk}.  Usually in random sequential adsorption
problems, it is convenient to study the complementary density of gaps
between domains which satisfies a {\it linear} rate equation.
Nevertheless, the IIA is exact in this case as no ``mixing'' of
domains due to diffusion occurs.  The final domain density is given by
$N(\infty)=\exp[-q/(q-1)]$. The systems quickly reaches a jamming
configuration where domain walls are isolated and immobile. Thus, no
coarsening occurs and the postulated scaling behavior does not
apply. Nevertheless, as will be shown below, the domain size and
number distributions and in particular their tails do resemble their
diffusive counterparts.

The length distribution $P_n$ can be found using normalized 
distribution $p_n=P_n/N$ which satisfies (for $n\geq 2$)

\begin{equation}
\label{pns1}
{dp_n\over dt}={p_1\over q-1}
\left[\sum_{i=1}^{n-2} p_ip_{n-1-i}+(q-2)p_{n-1}\right].
\end{equation}
This equation can be further simplified by introducing the modified 
time variable $T$, defined via $dT/dt=p_1$ implying $T=1-e^{-t}$. 
To solve Eq.~(\ref{pns1}) we introduce the generating functions 

\begin{equation}
\label{pzt1}
p(z,T)=\sum_{n=1}^\infty p_n(T)z^n,
\end{equation}
that satisfies

\begin{equation}
\label{pzT}
{dp(z,T)\over dT}={z\over q-1}\,[p(z,T)^2+(q-2)p(z,T)]-z. 
\end{equation}
Solving Eq.~(\ref{pzT}) subject to the monodisperse initial conditions 
$P_n(0)=\delta_{n,1}$, i.e., $p(z,T=0)=z$, we get 

\begin{equation}
\label{pzt2}
p(z,T)=1+q\left[{z+q-1\over z-1}\exp
\left(-{qzT\over q-1}\right)-1\right]^{-1}.
\end{equation}

Clearly, quantities such as the domain density and the domain length
distribution approach exponentially fast their limiting values.  We are
especially interested in these limiting values.  By expanding the
generating functions in powers of $z$ we get the limiting density of
short domains $p_n(\infty)=0$, ${q-2\over 2(q-1)}$, 
${q^2-3q+3\over 3(q-1)^2}$, and ${q^2(q-2)\over 8(q-3)^3}$ for $n=1,2,3,4$.

Similar to the behavior seen in the Glauber-Potts model, large 
domains are suppressed exponentially, 

\begin{equation}
\label{pns2}
P_n(t)\sim [\lambda(q,t)]^n \qquad n\gg 1. 
\end{equation} 
Here $\lambda(q,t)$ is equal to the inverse of the first simple pole
of the generating functions $p(z,t)$. This can be seen from the
$\sum_n \lambda^n z^n\propto (\lambda^{-1}-z)^{-1}$ .  Note that
$\lambda(q,t)$ vanishes when $t\to 0$ as implied by the initial
conditions.  When $t\to\infty$, $\lambda(q,t)\to \lambda_\infty(q)$.
In particular when $q\to\infty$, $\lambda_{\infty}(q)$ vanishes
according to $\lambda_{\infty}(q)\sim 1/(\ln q)$, indicating a faster
than exponential limiting behavior. Indeed, when $q=\infty$ the
generating functions $p(z,\infty)=1+(z-1)e^{z}$ gives an 
inverse factorial decay, $p_n(\infty)=(n-1)/n!\sim e^{-n\ln n}$. In the
complementary $q=1$ limit, the density vanishes, as was the case in 
the diffusive counterpart.

The domain number-size distribution can be obtained by generalizing 
Eq.~(\ref{pns})

\begin{eqnarray}
\label{pnms}
{dP_{n,m}\over dt} &=& {P_1\over (q-1)N^2}
\left[\sum_{i,j}P_{i,j}P_{n-1-i,m-j}
-2NP_{n,m}\right]\nonumber\\
&+&{(q-2)P_1\over (q-1)N}\left[P_{n-1,m}-P_{n,m}\right]-
\delta_{n,1}\delta_{m,1}P_1.
\end{eqnarray}
A solution using the generating functions technique is possible here
as well. However, this solution is too cumbersome, and we briefly
discuss its qualitative features.  There are two limiting cases.  When
$q=\infty$, the joint domain-number distribution simplifies to
$P_{n,m}=P_n\delta_{m,1}$.  When $q=2$, domains are always of odd
length and $P_{n,m}=P_{n}\delta_{m,(n+1)/2}$.  Hence, the domain
number distribution also decays exponentially, $Q_m(t)\sim
[\Lambda(q,t)]^n$.  Similar to the length distribution, the decay
constant $\Lambda(q)$ vanishes when $q\to\infty$.

In summary, although the restricted dynamics Potts model does not
exhibit coarsening or scale invariance, the number and size
distributions mimic their diffusive counterpart large $n$ and large
$q$ behavior.

\subsection{Ballistic Annihilation Model}

Consider a binary reaction process with particles moving ballistically
and annihilating upon collision.  Assuming a bimodal velocity
distribution, we set these velocities equal to $\pm 1$, without loss
of generality.  Identifying domain walls with particles, this
two-velocity ballistic annihilation process\cite{ef} is equivalent to
deterministic coarsening in a system with three equilibrium
states\cite{lpe,fk}.

The domain size distribution for this ballistic annihilation process
has  been investigated in\cite{fk}.  Here we want to compute the
domain survival probability $S(t)$.  There are actually four such
survival probabilities depending on the initial velocities of boundary
interfaces; we denote the corresponding survival probabilities by
$S_{++}(t), S_{+-}(t), S_{-+}(t)$, and $S_{--}(t)$.  Then the total
survival probability is just the sum $S(t)={1\over 4}
\left[S_{++}(t)+S_{+-}(t)+S_{-+}(t)+S_{--}(t)\right]$.  We need to
specify the initial conditions.  Let us assume that interfaces are
located according to the Poisson distribution with unit density.  For
such symmetric initial conditions we have $S_{++}(t)=S_{--}(t)$.  One
immediately gets the survival probability in the simplest case when
the interfaces move toward each other:

\begin{equation}
\label{s+-}
S_{+-}(t)=e^{-2t}.
\end{equation}
To compute the survival probability of parallel moving interfaces we
note that the probability $S_+(t)$ for a single right-moving interface
to survive is $S_+(t)=[S_{++}(t)+S_{+-}(t)]/2$.  Combining the known
result\cite{ef}

\begin{equation}
\label{exact}
S_+(t)=S_-(t)=e^{-2t}[I_0(2t)+I_1(2t)], 
\end{equation}
with Eq.~(\ref{s+-}) we arrive at

\begin{equation}
\label{s++}
S_{++}(t)=S_{--}(t)=e^{-2t}[2I_0(2t)+2I_1(2t)-1].
\end{equation}
Using the asymptotic behavior\cite{bo} 
$I_n(t)\simeq e^t/\sqrt{2\pi t}$ when $t\to\infty$,
we get $S_{++}(t)\simeq 2/\sqrt{\pi t}$.

We turn now to the more challenging problem, i.e. to computation of
the survival probability when the interfaces move away from each
other.  The final answer is relatively simple:

\begin{equation}
\label{s-+}
S_{-+}(t)=e^{-2t}[2I_1(2t)+4I_2(2t)+2I_3(2t)+1]
\end{equation}
so that the total survival probability reads

\begin{equation}
\label{total}
S(t)={1\over 2}\,e^{-2t}[2I_0(2t)+3I_1(2t)+2I_2(2t)+I_3(2t)].
\end{equation}
Eqs.~(\ref{s-+}) and (\ref{total}) imply an algebraic asymptotic
behavior, $S_{-+}(t)\simeq 4/\sqrt{\pi t}$ and $S(t)\simeq 2/\sqrt{\pi
t}$ for $t\to\infty$.  Thus, the domain decay exponent is $\psi=1/2$.

To derive Eqs.~(\ref{s-+}) and (\ref{total}) let us consider a 
sequence of interfaces starting from the right interface of our domain
as a random walk.  Namely, we set $W_0=0$, then define
$W_1=W_0+v_1=v_1$ where $v_1=1$ is the velocity of the right interface
of our domain.  We repeat this procedure so that $W_j=W_{j-1}+v_j$ and
we treat $W_j$ as the displacement of the random walk, started from
the origin, at the  $j^{\rm th}$ step.  When the displacement becomes
negative for the first time, the corresponding interface will move to
the left and will eventually destroy the domain.  Let we meet this
interface after $2N+1$ steps.  The corresponding probability $P_N$ is
readily determined by random walk methods\cite{feller}:

\begin{equation}
\label{fel}
P_N=2^{-2N}{(2N)!\over (N+1)!N!}.
\end{equation}
The same analysis applies to the left interface of the domain.  Thus
we have $2N+1$ interfaces to the right and $2M+1$ interfaces to the
left.  Our original domain survives till time $t$ if the distance
between the extreme interfaces is greater than $2t$.  In other words,
the interval of length $2t$ with the left boundary at the initial
location of the extreme left interface should contain $2N+2M$
interfaces at most.  The probability
of this event is 

\begin{eqnarray}
\label{unm}
U_{N+M}(t)&=&e^{-2t}\sum_{k=0}^{N+M}{(2t)^k\over k!}\nonumber\\
          &=&\int_{2t}^\infty du\,e^{-u}\,{u^{2N+2M}\over (2N+2M)!}.
\end{eqnarray}
The survival probability $S_{-+}(t)$ is now given by

\begin{equation}
\label{s-+t}
S_{-+}(t)=\sum_{N,M\geq 1}P_N P_M U_{N+M}(t).
\end{equation}
It proves convenient to expand the summation in (\ref{s-+t}) to
$N=0$ and $M=0$.  This gives

\begin{eqnarray}
\label{s1}
S_{-+}(t)&=&\sum_{N,M\geq 0}P_N P_M U_{N+M}(t)\nonumber\\
&-&2\sum_{N\geq 0}P_N U_{N}(t)+U_0(t).
\end{eqnarray}
The second sum in Eq.~(\ref{s1}) can be rewritten as

\begin{equation}
\label{s2}
\sum_{N=0}^\infty P_N U_{N}(t)=
\int_{2t}^\infty du\,e^{-u}\sum_{N=0}^\infty
P_N\,{u^{2N}\over (2N)!}.
\end{equation}
The sum on the right-hand side of Eq.~(\ref{s2}) is compressed 
into $\sum_{N\geq 0}P_N\,{u^{2N}/(2N)!}=2I_1(u)/u$,
and the resulting integral is 

\begin{equation}
\label{s3}
\int_{2t}^\infty du\,e^{-u}\,{I_1(u)\over u}
=e^{-2t}[I_0(2t)+I_1(2t)].
\end{equation}
Thus $S_{-+}(t)$ becomes

\begin{eqnarray}
\label{s4}
S_{-+}(t)&=&\int_{2t}^\infty du\,e^{-u}
\sum_{N,M\geq 0}P_N P_M\,{u^{2N+2M}\over (2N+2M)!}\nonumber\\
&-&4e^{-2t}[I_0(2t)+I_1(2t)]+e^{-2t}.
\end{eqnarray}
$S(t)$ is found from (\ref{s+-}), (\ref{s++}), and 
(\ref{s4}) to yield

\begin{equation}
\label{s5}
S(t)={1\over 4}\,\int_{2t}^\infty du\,e^{-u}
\sum_{N,M\geq 0}P_N P_M\,{u^{2N+2M}\over (2N+2M)!}.
\end{equation}
To perform the summation in Eq.~(\ref{s5}) we need the combinatorial
identity $\sum_{N+M=L}P_N P_M=4P_{L+1}$ which can be checked directly.
One can also establish this identity geometrically by noting that
$2^{2N}P_N=(2N)!/N!(N+1)!$ gives the number of random walks starting
at the origin and returning to the origin for the first time after
$2N+2$ steps\cite{feller}.  An appropriate counting of all such walks
of length $2L+4$ then leads to the above identity.  Making use of this
identity we reduce Eq.~(\ref{s5}) to

\begin{eqnarray}
\label{long}
S(t)&=&{1\over 4}\,\int_{2t}^\infty du\,e^{-u}
\sum_{L=0}^\infty \left({u\over 2}\right)^{2L}
{2(2L+1)\over L!(L+2)!}\nonumber\\
&=&2\int_{2t}^\infty du\,e^{-u}
\left[{I_1(u)\over u}-{3I_2(u)\over u^2}\right]\\
&=&{1\over 4}\,\int_{2t}^\infty du\,e^{-u}
\left[I_0(u)-I_4(u)\right].\nonumber
\end{eqnarray}
In deriving the second line we have used the definition of the
modified Bessel functions; the third line has been derived by applying
the identity \cite{bo} $I_{n-1}(u)-I_{n+1}(u)={2n\over u}\,I_n(u)$.
Computing now the integral in the last line of Eq.~(\ref{long}) we 
arrive at (\ref{total}).  This completes the proof of Eqs.~(\ref{s-+})
and (\ref{total}).

One can try to compute $Q_m(t)$, the domain number
density.  First we note that $Q_1(t)$ with specified
boundary velocities can be readily found: $Q_{+-}(t)=e^{-2t}$, while
other single-domain densities $Q_{++}(t)=Q_{--}(t)$ and $Q_{-+}(t)$
can be expressed via single-particle survival probabilities
$S_{\pm}(t)$:

\begin{equation}
\label{q}
Q_{++}(t)=S_{+}(t),  \quad 
Q_{-+}(t)=S_{-}(t)S_{+}(t).
\end{equation}
Then the total single-domain survival probability reads
$Q_1(t)=[Q_{++}(t)+Q_{+-}(t)+Q_{-+}(t)+Q_{--}(t)]/4$.  Asymptotically,
$Q_1(t)\simeq (4\pi t)^{-1/2}$ implying exponents $\delta=\psi=\nu=1/2$.
Moreover the persistence exponent $\theta=1$, and thus all exponents
are identical with those of the $q=\infty$ Potts model.  These two
models exhibit several other similarities\cite{lps}.  However, the
present deterministic model of coarsening is quite different in that
the number distribution $Q_m(t)$ is nontrivial. The determination of
this distribution is more involved and left for the future. 

\section{DISCUSSION AND SUMMARY}

Even in one dimension there are many interesting situations where the
above coarsening exponents are unknown.  The simplest case is a
diffusion equation which can describe coarsening in systems with
non-conserved order parameter\cite{bray}.  Recently, the persistence
characteristics for the diffusion equation process have been
investigated\cite{msbc} numerically and theoretically by an approach
close to the IIA.  Given the enormous role played by the diffusion
equation in science, surprisingly little is known about its underlying
coarsening process \cite{w}.

Another well known coarsening process is the 1D time-dependent
Ginzburg-Landau equation for a scalar non-conserved
order parameter\cite{bray}.  In this system, domains do not move and
the coarsening proceeds via flipping of the shortest domains.  The
minimum size grows logarithmically\cite{nk}, so it is convenient to
define the coarsening exponents in terms of the minimum size $L$
rather than time $t$.  This process is solvable in that the domain
size distribution, $P_n(L)=L^{-2}{\cal P}(n/L)$, is
known\cite{nk,bray}.  The same expression holds for the domain number
distribution.  Some coarsening exponents are simple, $\nu=1$ and
$\delta=\sigma=\infty$. In contrast, the persistence exponent,
$\theta\cong 0.1750758$\cite{dbg1}, is non-trivial.  This process 
resembles  the $q\to 1$ Potts model where $\psi=\theta$ as well.  

It would be interesting to extend of our work to coarsening systems
with {\em conserved} order parameter.  Besides the dynamical exponent
$z=3$ little is known even for the one-dimensional Ising model with
Kawasaki spin-exchange dynamics \cite{cor}. The coarsening exponents
$\psi$ and $\delta$ appear to be non-trivial for the Ising-Kawasaki
model\cite{ebnpk} as well. Another possible direction is to study the
coarsening exponents for the natural generalization of the ballistic
annihilation process, the $N$-species Lotka-Volterra
process\cite{lpe,tai}.

In summary, we introduced the domain size distribution and showed that
it obeys scaling and is characterized by two independent nontrivial
decay exponents.  The survival probabilities of a domain and an
unreacted domain are described by the exponents $\psi$ and $\delta$,
respectively.  Generally, these exponents obey $0\le\psi\le\theta$ and
$\psi\le\nu\le\delta$.  In the most examples the above inequalities
are strict; however, there are counter examples where $\psi=\theta$
and/or $\delta=\nu$.  For the 1D $T=0$ $q$-state Potts-Glauber model
we developed the IIA that predicts the correct qualitative behavior of
the domain size and number distributions and even reasonable estimates
for the decay exponents.  We also worked out the analytically
tractable limits of $q\to 1$ and $q\to\infty$.  It still remains,
however, to obtain the exact behavior for general $q$.  This might be
possible using the techniques used in studies of single-spin
persistence \cite{dhp,ms,cardy}.  In a static version of the Potts
model, an exact solution was presented and exponential decay of the
domain density still occurred.  It was also shown analytically that
the coarsening exponents in the solvable {\em deterministic} ballistic
annihilation model and the {\em stochastic} $q=\infty$ Potts model are
identical.

These results indicate that several nontrivial decay laws underlie the
evolution of elementary processes such as the non-equilibrium Ising
model.  These nontrivial exponents do not emerge naturally from
studies of traditional quantities such as spatiotemporal correlations.
It remains a challenge to find and obtain these underlying ``hidden''
exponents from a more systematic method.  It is also intriguing to
determine whether an entire hierarchy or a finite number of
independent decay modes are present in these systems.

\vspace{.2in}
\centerline{\bf ACKNOWLEDGMENTS}
\vskip .1in
\noindent
We thank Stefan Boettcher, Laurent Frachebourg and Sidney Redner for
stimulating interactions.  The work of EB was supported by DOE and the
work of PLK was supported by ARO (grant DAAH04-96-1-0114).

\end{multicols}

\end{document}